# Energy Controlled Edge Formation for Graphene Nano Ribbons

Deepika[1], T.J. Dhilip Kumar[2], Rakesh Kumar[1*]

[1] *Department of Physics, Indian Institute of Technology Ropar, Rupnagar 140001, INDIA*
[2] *Department of Chemistry, Indian Institute of Technology Ropar, Rupnagar 140001, INDIA*
[*] *email: rakesh@iitrpr.ac.in*

On the basis of first principles calculations, we report energy estimated to cut a graphene sheet into nanoribbons of armchair and zigzag configurations. Our calculations show that the energy required to cut a graphene sheet into zigzag configuration is higher than that of armchair configuration by an order of 0.174 eV. Thus, a control over the threshold energy might be helpful in designing an experiment "How to cut a graphene sheet into smooth edged armchair or zigzag configurations."

## INTRODUCTION

Graphene, one atom thick layer of carbon atoms arranged in a honeycomb lattice [1], was discovered in 2004 by Geim's group at University of Manchester, although its electronic band structure reported by P. R. Wallace, was known since 1947 [2]. The relativistic speed of charge carriers in graphene, as a consequence of its linear band structure, gives rise to extraordinary electronic properties like room temperature Quantum Hall effect [1], and the highest mobility of charge carriers (suspended graphene). In addition, its two dimensional nature, structural and mechanical stability down to nanoscale makes it an ideal material capable to replace silicon in electronic devices in upcoming years. The main problem encountered with graphene is its semimetallic nature. For electronic applications, it should have a band gap like semiconductors. By quantum confinement of charge carriers a band gap can be opened in graphene by cutting it into nanoribbons [3]. Theoretical band structure calculations for graphene nanoribbons (GNRs) showed that in addition to quantum confinement, smoothness of edge configuration plays an important role in predicting its semiconducting or metallic nature [4]. But even after a long time, since its discovery, it has not been possible to develop a simple and reliable technique to make nanoribbons of smooth edges.

In this paper, on the basis of first principles calculations, we report energy required to cut a graphene sheet into GNRs of armchair and zigzag edge configurations and predict that external parameter like temperature may be a controlling factor.

## COMPUTATIONAL DETAILS

We performed non-magnetic theoretical simulations based on first principles calculations using Density Functional Theory (DFT). For electronic calculations, Vienna *ab initio* Simulation Package (VASP) [5] was used. Plane wave basis sets was used to describe the valence electron orbitals, Generalized Gradient Approximation (GGA) for electrons exchange-correlations, and Projected-Augmented Wave (PAW) potentials for electron-core interactions. A vacuum layer of 10 Å was taken along the z-direction in a unit cell to avoid interlayer interactions [6]. Brillouin zone integration was done using Monkhorst-Pack for k-mesh sampling. All atoms in a unit cell were relaxed to a force tolerance of 0.001 eV/Å. Pre-analysis calculations were done to optimize the size of k-mesh and cutoff energy. A mesh of 25x25x1 was taken for graphene sheet and 25x1x1 for GNRs. Cut off energy of 300 eV was taken for both graphene and GNRs. After relaxation of atoms in the unit cell, its ground state energy was calculated. To estimate C-H bond formation energy, Gaussian 09 package was used with CCSD(T)/cc-pvtz basis sets [7].

## RESULTS AND DISCUSSION

To find the energy required to cut a graphene sheet into armchair or zigzag configurations, we considered a two dimensional periodic unit cell for graphene sheet and from the same unit cell we constructed one dimensional periodic unit cell for GNRs of armchair (AGNR) and zigzag (ZGNR) configurations accordingly by hydrogenation at the edges. Hydrogenation at the edges of GNRs was incorporated for the stability of dangling bonds. Therefore, actual ground state energy of the GNRs was determined by subtracting the corresponding hydrogenation energy at the edges. The hydrogenation energy (C-H bond formation energy) of 4.172 eV was estimated from methane gas molecule including zero point correction. In GNRs, we observed that the ground state energy of a unit cell depends upon its width as well as its edge configurations. Therefore, to compare the edge formation energy for armchair and zigzag configurations, we took nearly the same width for both the configurations of GNRs; avoiding the energy contribution due to quantum confinement. Width of the unit cell for graphene sheet was taken to be 2.84 nm x 2.83 nm, which corresponds to a width of 24-AGNR and 14-ZGNRs as shown in figure 1. The ground state energy for the unit cell of graphene sheet and the corresponding GNRs of armchair and zigzag configurations are shown in table 1.

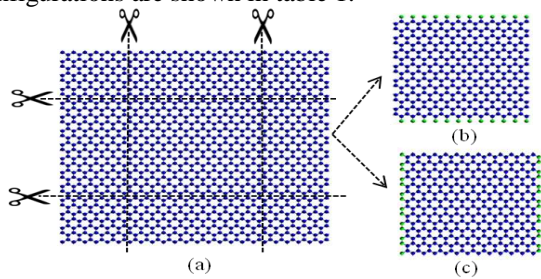

**FIGURE 1.** (a) Periodic unit cell of graphene sheet is shown by enclosed dotted lines. Periodic unit cell of Hydrogen terminated edges for 14-ZGNR and 24-AGNR are shown in (b) and (c), respectively. Number of dimer lines and zigzag chains along the periodic direction describes the width of the AGNRs and ZGNRs, respectively. Carbon atoms are shown in blue color and hydrogen atoms in green.

To find energy required for the formation of GNRs of the corresponding configurations from a graphene sheet, the actual ground state energy of the GNRs was subtracted from the ground state energy of the graphene sheet. Hence, total energy involved in the formation of the corresponding GNRs was determined by dissociation of associated C-C bonds in the graphene sheet. The dissociation energy per C-C bond for the formation of armchair and zigzag configurations from a graphene sheet was estimated to be 0.870 eV and 1.044 eV, respectively. The corresponding difference between the dissociation energy per C-C bond of armchair and zigzag configurations was estimated to be 0.174 eV.

**TABLE 1.** Ground state energy for periodic unit cell of graphene sheet, Hydrogen terminated armchair and zigzag configurations.

| Periodic unit cell | Width (nm) | Ground state energy (eV) |
|---|---|---|
| (14x24) graphene sheet | 2.84 x 2.83 | -3118.710 |
| 24-AGNR (length 2.84 nm) | 2.83 | -3211.168 |
| 14-ZGNR (length 2.83nm) | 2.84 | -3193.783 |

From the above results, we observed that (a) the energy required to form a zigzag configuration from a graphene sheet is higher than that of an armchair configuration, (b) the difference between edge formation energy of both the configurations corresponds to a temperature difference of the order of $10^3$ K. From these observations, it can be concluded that a control over the threshold energy using temperature as a controlling parameter might be helpful in designing an experiment "How to cut a graphene sheet into smooth edged armchair or zigzag configurations."

## ACKNOWLEDGMENT

Authors would like to thank whole team of Param Yuva II, CDAC-Pune for providing supercomputing facility, and IIT Ropar for providing research work support.

## REFERENCES

1. K. S. Novoselov *et al. Science* **315**, 1379 (2007).
2. P. R. Wallace, *Phys. Rev.* **71**, 622 (1947).
3. K. Nakada *et al. Phys. Rev. B.* **54**, 17954 (1996).
4. Y. W. Son *et al. Phys. Rev. Lett*. **97**, 216803 (2006).
5. G. Kresse and J. Furthmüller, *Comp. Mater. Sci.* **6,** 15 (1996).
6. De-en Jiang *et al. J. Chem. Phys.* **126**, 134701 (2007).
7. Gaussian 09, B.1, M.J. Frisch *et al.* (2010).